\documentclass[eqsecnum,showpacs,pra,superscriptaddress]{revtex4}

\usepackage{graphics}
\usepackage{amsmath}
\usepackage{amssymb}

\begin{document}

\title{Gaussian entanglement of symmetric 
two-mode Gaussian states}
\author{Paulina Marian} 
\affiliation{ Department of Chemistry, 
University of Bucharest, Boulevard Regina Elisabeta 4-12, 
R-030018 Bucharest, Romania}
\author{Tudor A. Marian}
\affiliation{ Department of Physics, 
University of Bucharest, P.O.Box MG-11,
R-077125 Bucharest-M\u{a}gurele, Romania}
\date{\today}
\begin{abstract}

A Gaussian degree of entanglement for a symmetric two-mode Gaussian
state can be defined as its distance to the set of all separable two-mode Gaussian
states. The principal property that enables us to evaluate both Bures distance and relative entropy between symmetric two-mode Gaussian
states is the diagonalization of their covariance matrices under the same beam-splitter transformation. The multiplicativity property of the Uhlmann  fidelity and the additivity of the relative entropy allow one to finally deal with a single-mode optimization problem in both cases. We find that only 
the Bures-distance Gaussian entanglement 
 is consistent with the exact entanglement of formation.
\end{abstract}
\pacs{03.67.-a; 42.50.Dv; 03.65.Ud; 03.67.Mn }
\maketitle

\section{Introduction}
\label{intro}

Intense recent work on  the entanglement of two-mode Gaussian states (TMGS's) 
was  stimulated by the  important result 
that preservation of the nonnegativity of their density matrix under 
partial transposition \cite{Peres} is not only a necessary, but also
a sufficient condition for their separability. 
Using the Sp $(2,\mathbb{R})  \times$ Sp $(2,\mathbb{R})$ 
invariant form of this criterion written by Simon \cite{Si1}, 
one can easily check whether a two-mode Gaussian state is separable 
or not \cite{PTH}. In spite of 
considerable effort in using some of the accepted measures of entanglement to the Gaussian-state case,
 the only {\em exact} evaluation at present
appears to be  the entanglement of formation 
(EoF) for  a symmetric TMGS \cite{G}. In this particular  case
the EoF proved to 
be a monotonous function of the smallest symplectic eigenvalue
 of the covariance matrix of
the partially transposed (PT) state. This eigenvalue will hereafter be denoted
 by $\tilde{\kappa}_-$.
 
A computable inseparability measure 
for an arbitrary bipartite state was  proposed by Vidal and 
Werner \cite{VW}  in terms of the sum of the negative 
eigenvalues of the PT- density matrix. For TMGS's, 
the absolute value of this sum, called  negativity  \cite{VW},
is an expression  depending  only on $\tilde{\kappa}_-$. It is thus  consistent to the EoF.
 As proved by Vidal and Werner, 
the negativity is an entanglement monotone.

The possibility of identifying the set of separable TMGS's \cite{Si1}
paved the way to the application of the distance-type proposal 
for quantifying entanglement made by Vedral and co-workers \cite{Ved}. 
A class of {\em 
 distance-type Gaussian measures 
of entanglement} was defined with respect to  only the set of Gaussian states. 
To the best of our knowledge, the first authors who  used and evaluated numerically a
 Gaussian measure of entanglement were Scheel and Welsch in Ref.\cite{Scheel}.
In our paper \cite{PTH1} co-authored with H. Scutaru, an explicit analytic
Gaussian amount of entanglement was calculated for
 two-mode squeezed thermal states (STS's) by using the
Bures distance. We then employed
the Gaussian  approximation for the entropic entanglement 
of a two-mode STS and evaluated it in the pure-state case. Comparison
to
 the von Neumann entropy of the subsystems (reduced modes)
 which was known to be the exact relative entropy of entanglement in the 
pure-state case \cite{Ved}, indicated an 
encouraging accuracy of the Gaussian approach.
Note that the STS's are important non-symmetric
TMGS that can be produced experimentally and are used  in the protocols 
for quantum teleportation.

Another Gaussian measure of entanglement, {\em the Gaussian entanglement of formation} (EoF) for an arbitrary TMGS 
was defined 
with respect to its optimal decomposition in Gaussian pure states \cite{W}.
 Analitically, the Gaussian EoF was evaluated for symmetric TMGS's 
and  was shown to coincide with the exact expression given in Ref.\cite{G}.
Following the prescription of Ref.\cite{W}, an evaluation of the Gaussian EoF
 for a STS was given in the paper  \cite{Jiang}. 
 In the general case an insightful formula was not yet written.   

One can notice that, for a symmetric TMGS, the amount of entanglement 
is fairly well described by monotonous functions 
  (negativity, EoF,
 and Gaussian EoF) depending on $\tilde{\kappa}_-$ only.
However, the situation is different for other special TMGS's.
 In the STS case, the Gaussian entanglement measured 
by Bures distance  \cite{PTH1} and
 the Gaussian EoF \cite{Jiang}  are found to be in agreement. They 
  are nicely
depending on the same parameter, the difference between the two-mode squeeze parameter $r$ and its 
value $r_s$ defining  the separability threshold. The parameter $r-r_s$   
 cannot be expressed  in terms of only $\tilde{\kappa}_-$.
Therefore, the negativity of a STS is not equivalent to the 
two Gaussian measures of its entanglement evaluated at present \cite{PTH1,Jiang}.
 A  similar disagreement
 between the Gaussian EoF and the negativity 
of the Gaussian states having extremal negativity at fixed global
 and local purities \cite{Aa} was recently noticed in Ref.\cite{AI}.
 
In this paper we  will compare 
two  distance-type Gaussian
entanglement measures to the exact EoF for a symmetric TMGS, checking thus on the validity of the Gaussian approach.  We recall in Section 2 several aspects of two-mode Gaussian states such as 
the diagonalization of the CM for a symmetric TMGS under
the beam-splitter transformation. 
   We define  a Gaussian degree of entanglement for a symmetric TMGS as its distance to the set of all separable TMGS.  As {\em distances} we employ the  Bures
distance in Sec.3 and the relative entropy in 
Sec.4.  
 In Sec. 3, by using the properties
 of the Uhlmann fidelity, we can restrict the reference set of 
all separable TMGS's to its subset of only symmetric TMGS's. Application 
of the beam-splitter transformation to both the given inseparable state 
and the set of symmetric separable  TMGS's enables us to evaluate and maximize
just a product of one-mode fidelities. Inspired by the results obtained in Sec.3, we define and calculate in Sec.4 an entropic Gaussian entanglement as the minimal relative entropy  between a symmetric TMGS 
and the set of all separable symmetric TMGS's.
 Our final conclusions are presented 
in Sec. 5.

\section{Two-mode Gaussian states}
 
An undisplaced TMGS is entirely specified 
by its covariance matrix (CM) denoted by ${\cal V}$ which determines 
the characteristic function of the state
\begin{eqnarray}
\chi_G(x)=\exp{\left(-\frac{1}{2}x^T {\cal V} x \right)},
\label{CF} 
\end{eqnarray}
with $x^T$ denoting a real row vector $(x_1\; x_2\; x_3\; x_4)$.The superscript $T$ stands for transpose.
${\cal V}$ is a  symmetric and positive $4\times 4$ 
 matrix which has the following block structure:
\begin{eqnarray}
{\cal V}=\left(\begin{array}{cc}{\cal V}_1&{\cal C}\\
 {\cal C}^T&{\cal V}_2 \end{array}\right).
\label{2.26} 
\end{eqnarray}
Here ${\cal V}_1$, ${\cal V}_2$, and ${\cal C}$ are $2\times 2$
matrices. Their entries are correlations of the canonical operators
$q_j=(a_j+a_j^{\dag})/{\sqrt{2}},\; p_j=(a_j-a_j^{\dag})/(\sqrt{2}i)$,
where $a_j $ and $a_j^{\dag}$,  $(j=1,2)$, are the amplitude operators 
of the modes. ${\cal V}_1$ and ${\cal V}_2$ denote the symmetric 
covariance matrices for the individual reduced one-mode STS's \cite{Ma},  
 while the matrix ${\cal C}$ contains the cross-correlations 
between modes. 
The Robertson-Schr\"odinger form of the uncertainty relations for the canonical
variables reads
\begin{equation}{\cal V}+\frac{i}{2}\Omega\geq 0.
\label{up}
\end{equation}
Here $\Omega$ is the $4\times 4$ fundamental symplectic block-diagonal matrix
\begin{eqnarray}
\Omega:=\left(\begin{array}{cc}J&0\\
 0&J \end{array}\right),\;\;J:=\left(\begin{array}{cc} 0&1\\-1&0\end{array}\right).
\label{Omega} 
\end{eqnarray}
From Eq.\ (\ref{up})
we have  \cite{Si1,PTH}
\begin{equation}
{\rm det} ({\cal V}+\frac{i}{2}\Omega)={\rm det}{\cal V}-\frac{1}{4}\left({\rm det}{\cal V}_1+
{\rm det}{\cal V}_2+2{\rm det}{\cal C}\right)+\frac{1}{16}\geq 0.
\label{Sp2} 
\end{equation}
A factorized form of the condition \ (\ref{Sp2}) in terms of the symplectic eigenvalues $\kappa_+$ and $\kappa_-$ of the CM,
 \begin{equation}
{\rm det} ({\cal V}+\frac{i}{2}\Omega)
=\left(\kappa_+^2 -
\frac{1}{4}\right)\left(\kappa_-^2-\frac{1}{4}\right)\geq 0,\label{fact1}
\end{equation}
shows that $\kappa_+\geq \kappa_- \geq 1/2.$

As stated by the separability criterion  \cite{Si1}, 
a TMGS is separable if and only if the uncertainty relation 
\ (\ref{Sp2}) is satisfied by the partially transpose state (PTS)
${\rho^{PT}}$ whose CM is hereafter denoted by $\tilde{\cal V}$. 
Hence the separability condition is
\begin{equation}
{\rm det} (\tilde{\cal V}+\frac{i}{2}\Omega)
={\rm det}{\cal V}-\frac{1}{4}\left({\rm det}{\cal V}_1+
{\rm det}{\cal V}_2+2|{\rm det}{\cal C}|\right)+\frac{1}{16}\geq 0.
\label{Siminv} 
\end{equation}
Equivalently, it can be written in terms of the smallest 
symplectic eigenvalue of $\tilde{\cal V}$ as $\tilde{\kappa}_-\geq 1/2.$
\subsection{Standard forms of the CM}

According to the important Lemma 1 in Ref.\cite{Duan}, 
 the $4\times 4$ covariance 
matrix of a Gaussian state may be cast into a standard form ${\cal V}^{(I)}$ by local symplectic
transformations such that  
the submatrices ${\cal V}_1$, ${\cal V}_2$ are multiples 
of the $2\times 2$ identity matrix ${\cal I}$ and ${\cal C}$ is diagonal.
We have

\begin{eqnarray}
{\cal V}_1=b_1{\cal I}\;\;\; {\cal V}_2=b_2{\cal I},\;\;\;
{\cal C}=\left(\begin{array}{cc} c&0\\
0&d\end{array}\right),\;\;\;\left(b_1\geq\frac{1}{2}\;,\;\;
b_2\geq\frac{1}{2} \right).
\label{s1} 
\end{eqnarray}
 An obvious  one-to-one correspondence can be found between the set
 of the four standard-form 
parameters $b_1, \; b_2, \;c, \;d$ appearing as entries in 
${\cal V}^{(I)}$ and the set of the 
  Sp$(2,\mathbb{R}) \times $ Sp$(2,\mathbb{R})$ 
 invariants ($\det {\cal V}_1, \; \det {\cal V}_2, \; \det {\cal C}$, 
and $\det {\cal V}$). 
According to Simon \cite{Si1}, entangled TMGS's should have a negative $d$ parameter.

Another important form of the CM achieved by local squeezing transformations
of the standard CM  ${\cal V}^{(I)}$ was discovered by Duan {\em et al.} 
\cite{Duan} and termed  {\em the 
standard form II}, hereafter denoted by ${\cal V}_{II}$. It describes a TMGS for 
which the separability and classicality conditions coincide. Generally,
the  classicality condition (existence of a well--behaved $P$ representation)
 is stronger than the separability one, Eq.\ (\ref{Siminv}). See our
paper \cite{PTH} for a more detailed analysis
 on this issue. It was proved in Ref.\cite{Duan} that the squeezing factors $v_1,v_2$ defining the standard form II  satisfy the algebraic 
system   
\begin{eqnarray}\frac{{b_1}(v_1^2-1)}{2 b_1-v_1}=\frac{{b_2}(v_2^2-1)
}{2 b_2-v_2},
\label{f4a}\end{eqnarray}
\begin{eqnarray}b_1b_2(v_1^2-1)(v_2^2-1)=(c v_1v_2-|d|)^2.
\label{f4b}\end{eqnarray} 
The solution of the system \ (\ref{f4a})--\ (\ref{f4b})
for an arbitrary TMGS
arises finally from a still unsolved eighth-order one-variable algebraic
 equation. However, it is possible to find $v_1,v_2$ in some particular cases.

\subsection{Symmetric TMGS's}
When having  
$\det{\cal V}_1= \det{\cal V}_2=b^2$ 
we are dealing with {\em symmetric} TMGS's. The standard parameters 
of the CM's for 
 symmetric TMGS's are
 denoted as
$b:=b_1=b_2,\;c>|d|,\; d=-|d|$.  
 The symplectic eigenvalues of the CM are found to be
\begin{equation}\kappa_+=\sqrt{(b-|d|)(b+c)},\;\;\kappa_-=\sqrt{(b+|d|)(b-c)}.\label{simp}
\end{equation}
Equations \ (\ref{f4a}) and \ (\ref{f4b}) can be solved for a symmetric TMGS.
We readily get the squeezed factors in the standard form II
\begin{equation}v_1=v_2=\sqrt{\frac{b-|d|}{b-c}}. \label{v12}\end{equation}
Equation \ (\ref{Siminv}) factorizes 
\begin{equation}{\rm det} (\tilde{\cal V}+\frac{i}{2}\Omega)=\left[(b-|d|)(b-c)-
\frac{1}{4}\right]\left[(b+|d|)(b+c)-\frac{1}{4}\right]\geq 0,
\end{equation}
leading to the  separability condition \cite{Duan}
\begin{equation}(b-|d|)(b-c)-\frac{1}{4}\geq 0. \label{c2}\end{equation}
Remark that \begin{equation}
\tilde{\kappa}_-=\sqrt{(b-|d|)(b-c)}\label{c-}\end{equation}
is the smallest symplectic 
eigenvalue of the CM for the PTS.

The most important property of the CM of a symmetric TMGS is its diagonalization under a beam-splitter transformation.  The possibility of using this nice property to evaluate a distance-type Gaussian  entanglement was first pointed out by
de Oliveira in Ref.\cite{O}. The optical effect of a lossless beam splitter is described by the {\em 
wave mixing operator} \cite{bonny,leo}
\begin{equation}
B(\theta, \phi)=\exp{\left[-\frac{\theta}{2}(
{\rm e}^{i\phi}a_1^{\dag}a_2-{\rm e}^{-i\phi}a_1a_2^{\dag})\right]}
 \label{bs1}\end{equation}
with $\theta\in[0,\pi],\;\;\;\phi\in(-\pi,\pi).$
Transformation of an arbitrary CM is governed by a
 $4\times4$ symplectic and orthogonal 
matrix
${ M}(\theta,\phi)\in{\rm S O(4)}\cap {\rm Sp}(4,\mathbb{R})$
\begin{equation}\tilde{\cal V}= { M}^T {\cal V} { M}.
\label{bs2}\end{equation}
The explicit form of ${ M}(\theta,\phi)$ is given in Refs.\cite{bonny,leo}.
The CM of a symmetric state having equal local squeezing factors ($u=u_1=u_2$)   is diagonalized by the transformation 
\ (\ref{bs1})
having the angles $\phi=0$ and $\theta=\pi/2$. We obtain in a straightforward manner
\begin{eqnarray}\tilde{\cal V}(u,u)={\rm diag}[(b+c)u,(b-|d|)/u,(b-c)u,
(b+|d|)/u]. \label{di0}\end{eqnarray}
In the particular case of symmetric TMGS's having the CM's
 in the standard form II
  we get
\begin{eqnarray}\tilde{\cal V}^{(II)}
={\rm diag}\left[(b+c)\sqrt{
\frac{b-|d|}{b-c}}
,\tilde{\kappa}_-,\tilde{\kappa}_-,(b+|d|)\sqrt{\frac{b-c}{b-|d|}}\right]. 
\label{di1}
\end{eqnarray}

\section{Gaussian entanglement by Bures metric}
Vedral and co-workers \cite{Ved}
 characterized the degree of inseparability of any bipartite 
state by its distance to the set 
of 
 all separable states of the given system. Although the 
distance-type definition is an ideal measure 
of inseparability, one is usually forced to modify it by restricting
the set of all separable states to a  relevant one
identified by a separability criterion. For the continuous-variable 
two-mode systems, a separability 
criterion was proved only for TMGS's \cite{Si1,Duan}. We find thus natural
to use the separable TMGS's as reference set when defining an 
entanglement measure for a symmetric TMGS. All the states sharing the same local symplectic invariants have the same entanglement. For later convenience, we choose to evaluate the entanglement of a symmetric TMGS $\rho_s$ whose CM is in the standard form II.  Its parameters are denoted by $b,c,d=-|d|$ and the standard-form II squeezing factors by $v_1=v_2= \sqrt{(b-|d|)/(b-c)}$.
Among the 
defined distances \cite{F} we concentrate now on those 
providing the best distinguishability of quantum states \cite{PTH1}. From this 
point of view, the strongest candidates  are the Bures distance \cite{Bures} and  the relative entropy 
\cite{Weh,Ved}. 
We give here a short account of the results on considering the Bures metric as a measure of entanglement for symmetric Gaussian states recently obtained in our paper \cite{PT07}. Recall that the  Bures distance $d_{B}(\rho,\rho^{\prime})$
 between the density operators ${\rho}$ and ${\rho^{\prime}}$ 
acting on a Hilbert space ${\cal H}_{A}$ originally introduced on mathematical grounds \cite{Bures} was then  written by Uhlmann \cite{Uhl} as
\begin{equation}
d_{B}(\rho,{\rho}^{\prime}):=[2-2\sqrt{{\cal F}(\rho,{\rho}^{\prime})}]^{1/2}. 
\label{db}
\end{equation}
In Eq.\ (\ref{db}), 
${\cal F}(\rho,	{\rho}^{\prime})$ is the Uhlmann {\it fidelity} \cite{Uhl,Jo} 
of the two states. 
Uhlmann also derived an intrinsic formula of the fidelity \cite{Uhl}:
\begin{eqnarray}
{\cal F}(\rho, {\rho}^{\prime})=\left\{{\rm Tr}[(\sqrt{\rho}{\rho}^{\prime}\sqrt{\rho})
^{1/2}]\right\}^2. \label{3} 
\end{eqnarray}
Following \cite{Ved} we define the Bures-metric entanglement of the symmetric TMGS $\rho_s$
\begin{equation}
E_B(\rho_s):=\min_{{\rho}^{\prime} \in {\cal D}_0^{sep}}\frac{1}{2} 
d^2_B(\rho_s, {\rho}^{\prime})=1-\max_{{\rho}^{\prime} \in {\cal D}_0^{sep}}
\sqrt{{\cal F}(\rho_s, {\rho}^{\prime}}). 
\label{set} 
\end{equation}
In Eq.\ (\ref{set}) we have introduced 
the set ${\cal D}_0^{sep}$
of {\em all separable scaled standard TMGS} which is included in
the set of all separable TMGS. The states belonging to the set ${\cal D}_0^{sep}$ have their CM's of the type
\begin{eqnarray}{\cal V}^{\prime}(u^{\prime}_1,u^{\prime}_2)  =\left(\begin{array}{cccc}b^{\prime}_1 u^{\prime}_1 &0&
c^{\prime}\sqrt{u^{\prime}_1u^{\prime}_2} &0\\
0& b^{\prime}_1/{u^{\prime}_1}&0&d^{\prime}/{\sqrt{u^{\prime}_1 u^{\prime}_2}}\\ c^{\prime}\sqrt{u^{\prime}_1u^{\prime}_2}  &0&b^{\prime}_2 u^{\prime}_2&0\\0
&d^{\prime}/\sqrt{u^{\prime}_1u^{\prime}_2}&0&b^{\prime}_2/{u^{\prime}_2 }\end{array}\right),\;\;(b^{\prime}_1\geq 1/2,\;
 b^{\prime}_2\geq 1/2).\label{tri}
\end{eqnarray}  
Our task is to maximize the fidelity between the entangled symmetric
TMGS  $\rho_s$ 
and a state ${\rho}^{\prime}\in{\cal D}_0^{sep}$. 
As discussed in our paper 
\cite{PTH1} the closest separable state,
 say ${\rho^{\prime\prime}}$, has the property
\begin{equation}
\tilde{\kappa}_-^{\prime\prime}=1/2.
\label{s}\end{equation}

Among the remarkable general properties  of the fidelity listed and largely discussed in
Refs.\cite{Uhl,Jo,F}, the following two ones proved to be especially important to our problem:
 \begin{description}
\item  {\bf P1.}   ${\cal F}(U \rho U^{\dag}, U {\rho}^{\prime} U^{\dag})=
{\cal F}(\rho, 	\rho^{\prime}),\;\;\;$ (invariance under unitary transformations $U$).
\item {\bf P2.}  ${\cal F}(\rho_1 \otimes \rho_2, \rho_1^{\prime} \otimes {\rho}_2^{\prime})=
{\cal F}(\rho_1, \rho_1^{\prime}){\cal F}(\rho_2, \rho_2^{\prime}),\;\;\;$ 
(multiplicativity).
\end{description}
In our paper \cite{PT07}, we have considerably 
simplified the minimization  procedure required  by Eq.\ (\ref{set}) by showing that the closest separable scaled standard state ${\rho^{\prime\prime}}$ to a given symmetric scaled 
standard state having equal local squeeze factors  $u_1=u_2=u$  is  
a similar symmetric scaled standard state observing the threshold 
condition \ (\ref{s}).
 Therefore, the amount of Gaussian entanglement for a symmetric
 TMGS can be calculated in a simpler way, 
because the separable reference set  ${\cal D}_0^{sep}$ used in Eq.\ (\ref{set})
is in fact 
restricted to the set ${\cal D}_s^{sep}$ of 
symmetric scaled standard states.  We have then used the property {\bf P1.} of the fidelity with respect to the beam-splitter transformation \ (\ref{bs1}) at the angles $\phi=0$ and $\theta=\pi/2$. The CM's of the given state $\rho_s$ and any equally scaled symmetric state ${\rho}^{\prime}\in{\cal D}_s^{sep}$ became 
 diagonal. The multiplicativity property {\bf P2.} allowed us to
 reduce the evaluation
of fidelity  to a single-mode problem.  The maximal fidelity was finally obtained in an elegant manner due to our choice for the given state $\rho_s$ (namely the symmetric TMGS having the CM in the standard form II): 
\begin{eqnarray}
\max_{{\rho}^{\prime} \in {\cal D}_0^{sep}}
{\cal F}(\rho_s, {\rho}^{\prime})
=\frac{2 \tilde{\kappa}_-}{\left(\tilde{\kappa}_-+1/2\right)^2}
\label{maxf}\end{eqnarray}
 We found that the Gaussian degree of entanglement 
measured by the Bures distance,
\begin{equation}
E_B(\rho_s)=\frac{(\sqrt{2\tilde{\kappa}_-} -1)^2}{2\tilde{\kappa}_-+1}
,\;\;\tilde{\kappa}_-<1/2,\label{fin}
\end{equation}
depends only on the smallest symplectic eingenvalue $\tilde{\kappa}_-$
 of the covariance matrix of the PTS. It is thus
in agreement with the exact expression of the entanglement of formation 
for symmetric
TMGS's.

\section{ Gaussian relative entropy of entanglement}
The relative entropy of a state $\rho^{\prime} $ with respect to the state $\rho$ is defined as 
\begin{equation}
S(\rho^{\prime}/\rho):={\rm Tr}\left[\rho(\ln \rho-\ln \rho^{\prime})\right].
\label{1}\end{equation}
 It is evident that the relative entropy is not a true metric because it lacks for
 symmetry. Among the important properties of the relative entropy proved and discussed 
in the classic paper of Wehrl \cite{Weh} and the more recent ones
 of Vedral {\em et. al.} \cite{Ved}, we shall use here the following ones:
\begin{description}
\item  {\bf $\Pi 1$}:    $S(\rho^{\prime}/\rho)=S(U\rho^{\prime} U^{\dag}/U\rho U^{\dag}),\;\;\;$ (invariance under unitary transformations $U$)
\item {\bf $\Pi 2$}:   $
 S(\rho^{\prime}_1\otimes\rho^{\prime}_2/\rho_1\otimes\rho_2)=
S(\rho^{\prime}_1/\rho_1) +S(\rho^{\prime}_2/\rho_2).\;\;\;$ (additivity)
\end{description}

\subsection{Defining Gaussian relative entropy of entanglement}
 In Ref.\cite{Ved}, the minimal relative
entropy between a  state  of a two-component system and the set of all separable
states, now called the relative entropy of entanglement,
 was proved to be a good measure of entanglement. The minimization process
was realized in the important case of the pure states. For mixed ones no 
exact result could be found so far. 
In order to perform a comparison to the Bures-metric entanglement, we consider the same reference set ${\cal D}_s^{sep}$ of separable states and the same given entangled state  $\rho_s$ and
define  the Gaussian relative entropy of entanglement 
 \begin{equation}
E_S(\rho_s):=\min_{{\rho}^{\prime} \in {\cal D}_s^{sep}}S({\rho}^{\prime}/\rho_s). \label{qs} \end{equation}
Definition \ (\ref{qs}) allows us to use the simultaneous diagonalization of the CM's under the beam-splitter transformation at the angles $\phi=0$ and $\theta=\pi/2$ as a consequence  of the property {\bf $\Pi 1$}. The  transformed state of $\rho_s$ will be denoted by $\tilde{\rho}_s$ and  has the diagonal CM written as 
Eq.\ (\ref{di1}). The transformation of an arbitrary state ${\rho}^{\prime} \in {\cal D}_s^{sep}$ leads us to the state $\tilde{\rho}^{\prime}$ which is described by the  CM
\begin{eqnarray}{\cal V}_{\tilde{\rho}^{\prime}}
={\rm diag}\left[2 (b^{\prime}+c^{\prime})
(b^{\prime}-|d^{\prime}|),
1/2,1/2,2 (b^{\prime}+|d^{\prime}|)(b^{\prime}-c^{\prime})\right], 
\label{di2}
\end{eqnarray}
where the separability threshold condition \ (\ref{s}) was inserted.
Equation \ (\ref{qs}) becomes
 \begin{equation}
E_S(\rho_s)=E_S(\tilde{\rho}_s)=\min_{\tilde{\rho}^{\prime} \in {\cal D}_s^{sep}}S(\tilde{\rho}^{\prime}/\tilde{\rho}_s). \label{qs1} \end{equation}
A diagonal $4\times 4$ CM describes in fact a product--state. Let us denote by $\tilde{\rho}_1$ and $\tilde{\rho}_2$ the reduced one-mode states of $\tilde{\rho}_s$. 
According to Eq.\ (\ref{di1}), the  CM's of the states $\tilde{\rho}_1$ and $\tilde{\rho}_2$ are, via Eqs.\ (\ref{simp}) and \ (\ref{c-}),
\begin{eqnarray}{\cal V}_{\tilde{\rho}_1}=\left(\begin{array}{cc}\frac{\kappa_+^2}{\tilde{\kappa}_-} &0\\
0&\tilde{\kappa}_-\end{array}\right),\;\;{\cal V}_{\tilde{\rho}_2}=
\left(\begin{array}{cc}
\tilde{\kappa}_- &0\\
0&\frac{\kappa_-^2}{\tilde{\kappa}_-}\end{array}\right).
\label{2.2a} \end{eqnarray}

Equation \ (\ref{2.2a})  nicely depends on both symplectic eigenvalues of the CM  and also on the smallest symplectic eigenvalue of the partially transposed CM.
Similarly, the state $\tilde{\rho}^{\prime}$ has the structure
$\tilde{\rho}^{\prime}= 
\tilde{\rho}^{\prime}_1\otimes\tilde{\rho}^{\prime}_2$ with
\begin{eqnarray}{\cal V}_{\tilde{\rho}_1^{\prime}}=\left(\begin{array}{cc}2
(\kappa_+^{\prime})^2 &0\\
0&\frac{1}{2}\end{array}\right),\;\;{\cal V}_{\tilde{\rho}_2^{\prime}}=\left(\begin{array}{cc}
 \frac{1}{2}&0\\
0&2(\kappa_-^{\prime})^2\end{array}\right).\label{2.2b} \end{eqnarray}

We apply now the additivity property {\bf $\pi 2$} of the relative entropy 
and get the interesting result 
\begin{eqnarray}
S(\tilde{\rho}^{\prime}/\tilde{\rho}_s)=S(\tilde{\rho}^{\prime}_1
/\tilde{\rho}_1)+S(\tilde{\rho}_2^{\prime}/\tilde{\rho}_2).\label{ad1}\end{eqnarray}
Therefore, as in the case of Bures-metric entanglement discussed in Sec.3,
evaluation of the entropic Gaussian entanglement of symmetric TMGS's is a one-mode problem. We now take advantage of having previously derived a general formula for the relative entropy between two one-mode Gaussian states. The derivation was rigorously performed in our paper \cite{PTH04} co-authored with H. Scutaru in order to define an entropic degree of nonclassicality for one-mode Gaussian states.

\subsection{Evaluating Gaussian relative entropy of entanglement}

 We adapt Eq.(A 14) from the Appendix of Ref.\cite{PTH04} to the present case of two undisplaced one-mode states having diagonal $2\times 2$ CM's of the type
\begin{eqnarray}
{\cal V}:=\left(\begin{array}{cc}\sigma_{qq} &0\\
0&\sigma_{pp}\end{array}\right),\;\;\;{\cal V}^{\prime}:=\left(\begin{array}{cc}\sigma^{\prime}_{qq} &0\\
0&\sigma^{\prime}_{pp}\end{array}\right).
\label{2.22a} \end{eqnarray}
In Eq.\ (\ref{2.22a}), the entries of the CM's are   expectation values of the canonical operators such as $\sigma_{pp}=\langle p^2\rangle$. 
We readily get
\begin{eqnarray} S(\rho^{\prime}/\rho)&=&-S_N(\rho)+
\frac{1}{2}\ln{\left[\sqrt{\det{\cal V}^{\prime}}+\frac{1}{2}\right]}
\left[1+\frac{\sigma_{qq}\sigma^{\prime}_{pp}+\sigma_{pp}
\sigma^{\prime}_{qq}}{\sqrt{\det{\cal V}^{\prime}}}\right]\nonumber\\&&
+\frac{1}{2}\ln{\left[\sqrt{\det{\cal V}^{\prime}}-\frac{1}{2}\right]}
\left[1-\frac{\sigma_{qq}\sigma^{\prime}_{pp}+\sigma_{pp}
\sigma^{\prime}_{qq}}{\sqrt{\det{\cal V}^{\prime}}}\right]
\label{en1}.\end{eqnarray}
Here we have introduced the von Neumann entropy $S_N(\rho):=-{\rm Tr}(\rho\ln \rho)$.
For a one-mode Gaussian state we have \cite{PTH04}
\begin{eqnarray} S_N(\rho)=\left(\sqrt{\det {\cal V}}+1/2\right)\ln{\left(\sqrt{\det{\cal V}}+1/2\right)}-\left(\sqrt{\det{\cal V}}-1/2\right)\ln{\left(\sqrt{\det{\cal V}}-1/2\right)}.\label{vne}\end{eqnarray}
 By using Eq.\ (\ref{en1}) via Eqs.\ (\ref{2.2a}) and \ (\ref{2.2a}), we have for the two-mode relative entropy, Eq.\ (\ref{ad1}),
\begin{eqnarray}
S(\tilde{\rho}^{\prime}/\tilde{\rho}_s)&&=-S_N(\rho_1)-S_N(\rho_2)\nonumber\\&&
+\frac{1}{2}\left\{\ln{\left[x_1+\frac{1}{2}\right]}\left[1+\frac{\kappa_+^2+4x_1^2(\tilde{\kappa}_-)^2}{2x_1\tilde{\kappa}_-}\right]+\ln{\left[x_1-\frac{1}{2}\right]}\left[1-\frac{\kappa_+^2+4x_1^2(\tilde{\kappa}_-)^2}{2x_1\tilde{\kappa}_-}\right]\right\}\nonumber\\&&+\frac{1}{2}\left\{\ln{\left[x_2+\frac{1}{2}\right]}\left[1+\frac{\kappa_-^2+4x_2^2(\tilde{\kappa}_-)^2}{2x_2\tilde{\kappa}_-}\right]+\ln{\left[x_2-\frac{1}{2}\right]}\left[1-\frac{\kappa_-^2+4x_2^2(\tilde{\kappa}_-)^2}{2x_2\tilde{\kappa}_-}\right]\right\}.
\label{relen}\end{eqnarray}
In Eq.\ (\ref{relen}) we have denoted
\begin{eqnarray}x_1=\kappa_+^{\prime},\;\;x_2=\kappa_-^{\prime},\label{r}\end{eqnarray}
the symplectic eigenvalues of the separable state $\tilde{\rho^{\prime}}$. The minimization of the relative entropy required by the definition \ (\ref{qs}) will be performed with respect to the variables $x_1$ and $x_2$ which are independent and separate in the expression \ (\ref{relen}) as a consequence of the additivity property \ (\ref{ad1}). Therefore, we can  formulate the following statement: 

{\em The Gaussian relative entropy of entanglement of a TMGS which can be unitarily transformed in a product--state equals the sum of the nonclassicality degrees of the transformed one-mode reductions}: 
\begin{equation}
E_S(\rho_s)=Q_S(\tilde{\rho}_1)+Q_S(\tilde{\rho}_2),\end{equation}
where  $Q_S(\rho)$ is the entropic nonclassicality degree of the one-mode Gaussian
 state $\rho$ defined and evaluated in our paper \cite{PTH04}. Note that
 in order to find the absolute minimum of the relative entropy one has to solve a transcendental equation, which does not have an exact analytic solution, but can be analyzed graphically as in Ref.\cite{PTH04}. Equation \ (\ref{relen}) displays a dependence on the symplectic eigenvalues $\kappa_-,\kappa_+$  and the smallest symplectic eigenvalue $\tilde{\kappa}_-$ of the PTS. The transcendental nature of the minimization condition suggests  that the Gaussian relative entropy of entanglement will not depend  only on $\tilde{\kappa}_-$.
Note finally that in Ref.\cite{Chen}, a general formula for the relative entropy of TMGS's was given by exploiting the exponential form of their density operators.  
 
\section{Conclusions}
The symmetric TMGS is  the only continuous-variable state for which an {\em exact} measure of entanglement is evaluated at present. Its EoF was found as depending only on  $\tilde{\kappa}_-$ \cite{G}. In this paper we have analyzed two distance-type Gaussian degrees of entanglement for a symmetric TMGS in order to compare our results to the exact EoF and check on the validity of the Gaussian approach. The principal property that enabled us to evaluate both Bures distance and relative entropy between two symmetric TMGS is the possibility of diagonalizing their CM's  under the same beam-splitter transformation. Calculation was simplified by
 considering the given state with its CM in the standard form II. Notice that
 this form of the CM is involved in giving an inseparability criterion
for TMGS \cite{Duan}. The remarkable multiplicativity property of Uhlmann fidelity and the additivity of the relative entropy allowed us to deal with a single-mode optimization problem in both cases.
Our result for the Gaussian degree of entanglement measured by Bures distance
depends only on the smallest symplectic eingenvalue
 of the covariance matrix of the PTS. Thus, it is 
in agreement with the exact EoF found in Ref.\cite{G}
 and enforces our previous 
idea 
\cite{PTH1}
 that the Bures distance is a reliable 
measure of entanglement  in  the Gaussian approximation. 
\section*{Acknowledgments}
This work was supported by the Romanian 
Ministry of Education and Research 
through the grant  CEEX 05-D11-68/2005 for the University of Bucharest.

\end{document}